\pdfoutput=1
\documentclass[a4paper,twoside]{article}
\usepackage{flushend}

\usepackage{url}
\usepackage{array}
\usepackage{graphicx}
\usepackage{algorithm,algpseudocode}
\usepackage{setspace}
\let\Algorithm\algorithm
\renewcommand\algorithm[1][]{\Algorithm[#1]\setstretch{1.2}}

\usepackage{subfigure}
\usepackage{calc}
\usepackage{amssymb}
\usepackage{amstext}
\usepackage{amsmath}
\usepackage{amsthm}
\usepackage{multicol}
\usepackage{pslatex}
\usepackage{apalike}
\usepackage{SCITEPRESS}     

\begin{document}

\title{Using a Predator-Prey Model to Explain Variations of Cloud Spot Price}

\author{\authorname{Zheng Li\sup{1}, William T{\"{a}}rneberg\sup{1}, Maria Kihl\sup{1} and Anders Robertsson\sup{2}}
\affiliation{\sup{1}Department of Electrical and Information Technology, Faculty of Engineering, Lund University, Lund, Sweden}
\affiliation{\sup{2}Department of Automatic Control, Faculty of Engineering, Lund University, Lund, Sweden}
\email{\{Zheng.Li, William.Tarneberg, Maria.Kihl\}@eit.lth.se, Anders.Robertsson@control.lth.se}
}


\abstract{The spot pricing scheme has been considered to be resource-efficient for providers and cost-effective for consumers in the Cloud market. Nevertheless, unlike the static and straightforward strategies of trading on-demand and reserved Cloud services, the market-driven mechanism for trading spot service would be complicated for both implementation and understanding. The largely invisible market activities and their complex interactions could especially make Cloud consumers hesitate to enter the spot market. To reduce the complexity in understanding the Cloud spot market, we decided to reveal the backend information behind spot price variations. Inspired by the methodology of reverse engineering, we developed a Predator-Prey model that can simulate the interactions between demand and resource based on the visible spot price traces. The simulation results have shown some basic regular patterns of market activities with respect to Amazon's spot instance type m3.large. Although the findings of this study need further validation by using practical data, our work essentially suggests a promising approach (i.e.~using a Predator-Prey model) to investigate spot market activities.}

\onecolumn \maketitle \normalsize \vfill

\section{\uppercase{Introduction}}
\label{sec:introduction}
The de facto Cloud market employs three types of pricing schemes for trading on-demand service, reserved service, and spot service respectively. With the on-demand service pricing scheme, Cloud consumers pay a fixed cost per service unit on an hourly basis for necessary on-demand resources, and an analogy of this pricing scheme is paying per view from a video on demand (VOD) service. With the reserved service pricing scheme, Cloud consumers pay an upfront fixed fee to ensure discounted hourly pricing for a long-term commitment of service availability, and an analogy of this pricing scheme is signing a two-year subscription of mobile service to receive cheaper data plans with a free phone. These two types of static pricing schemes both imply a straightforward demand-resource relationship when consuming Cloud services. In contrast, the spot pricing scheme depends on potentially complicated interactions between consumer demand and Cloud resource. As specified by Amazon \cite{Amazon_2015}, the price of Cloud spot service could frequently vary driven by a market mechanism. An analogy of this pricing scheme is the dynamic
pricing in the electricity distribution industry. Behind the price variations, technically, a Cloud spot service continuously evaluates its available resources and monitors the coming demands, and then dynamically sets spot prices to target predefined goals like revenue maximization or utility efficiency.

Given the generally low utilization of Cloud resources \cite{Delimitrou_Kozyrakis_2014}, although the two static pricing schemes are dominant trading strategies in the current Cloud market \cite{Al-Roomi_2013,Xu_Li_2013}, spot pricing has been considered to be a significant supplement for building a full-fledged market economy for the Cloud ecosystem \cite{Abhishek_Kash_2012}. However, it seems that both providers and consumers are still hesitating to enter the Cloud spot market. In fact, considering the hard-to-predict and dynamic interactions between demand and resource, the market-driven mechanism for pricing
spot service would require more effort and managerial
overheads for Cloud providers to implement, and also result in psychological difficulties for Cloud consumers to understand and employ \cite{Xu_Li_2013}. As can be seen, the overwhelming majority of the existing
Cloud providers have not employed the spot pricing scheme
yet \cite{Zaman_Grosu_2011}, and the only spot service provider Amazon 
is still using contests to encourage more spot applications
\cite{Amazon_2015}.

Therefore, being aware of the dynamic demands and resources would be significantly helpful and useful for both Cloud providers and consumers to join the spot market. Unfortunately, the backend details behind changing spot prices are invisible for most of the market participants, and little work has focused on the interactions between demand and resource. 
Following the methodology of reverse engineering, we tried to reveal the invisible knowledge from the tangible spot prices. By imaging the Cloud spot demand and resource as two species, i.e.~predator and prey respectively, we developed a Predator-Prey model to investigate the demand-resource interactions. Based on Amazon's historical spot prices, the simulation shows that our Predator-Prey model is conceptually functional, although the revealed information needs further validation in practice. This paper introduces our developed Predator-Prey model and uses its simulation to try explaining the variations of Cloud spot price. 

The contribution of our work is mainly twofold. Firstly, this work suggests a promising approach (i.e.~using a Predator-Prey model) to investigate spot market activities. To our best knowledge, this is the first study that tries to visualize the interactions between demand and resource in the Cloud spot market. Although the current version of this Predator-Prey model might still suffer from simple assumptions, the logic of the whole work can be reused and refined by others. Secondly, by using this Predator-Prey model, our simulation has identified some basic regular patterns of market activities with respect to Amazon's spot instance type m3.large\footnote{\url{http://aws.amazon.com/ec2/instance-types/}}. For example, spot resources could be accumulated relatively slowly, while being saturated quickly after reaching particular amounts. Such a phenomenon of sharp drops of spot resources might indicate ``herd behaviors" of spot demands. Although this simulation finding could not be practically assured at this current stage, it have provided us a hypothesis to be tested in the future. 

The remainder of this paper is organized as follows. Section \ref{sec:relatedwork} summarizes relevant studies that have modeled demand and resource of Cloud spot services by roughly classifying them into two types. Section \ref{sec:ppmodel} elaborates our development details of the Predator-Prey model of Cloud spot demands and resources. By using Amazon's spot price trace, Section \ref{sec:simulation} describes our simulation work that reveals basic information behind the changing spot prices. Conclusions and some future work are discussed in Section \ref{sec:conclusion}. 

\section{\uppercase{Related Work}}
\label{sec:relatedwork}
Although the backend details behind spot prices are usually uncertain and even unknown, the demand information and resource information are fundamentally crucial for investigating various problems ranging from service fault tolerance (from the consumer's perspective) to revenue maximization (from the provider's perspective). Therefore, researchers and practitioners have employed different techniques/assumptions to model the spot service demand and resource to facilitate their studies. The existing study approaches can be roughly classified into three categories, as specified below. 

(1) The first type of studies focuses on Cloud demand and resource separately.

\textit{\textbf{Demand model-relevant studies:}}

When it comes to modeling demands, a common practice was to replay recorded workload traces (e.g., \cite{Shi_Zhang_2014}) or to run particular applications (e.g., \cite{Chohan_Castillo_2010}). However, this type of practices would fail in emulating highly variable and dynamic workload profiles and demand scenarios \cite{Kistowski_Herbst_2014}. To address this limit, a simple case was to treat demand as an independent uncertainty parameter that contains the possible amounts of server-hours required by an application \cite{Chaisiri_Kaewpuang_2011}. In sophisticated cases, demand was imagined as job arrivals. For example, Abhishek et al.~\cite{Abhishek_Kash_2012} assumed that jobs arrived sequentially according to a stationary stochastic process with independent inter-arrival time, while Mazzucco and Dumas \cite{Mazzucco_Dumas_2011} assumed that jobs entered the spot service according to an independent Poisson process with a particular rate $\lambda$. 
Nevertheless, it is notable that these studies did not consider the influence of dynamic demands on spot prices in their modeling work.

\textit{\textbf{Resource model-relevant studies:}}

As for modeling resource, some authors simply assumed that the Cloud infrastructure could provide infinite/unbounded spot resource \cite{Chaisiri_Kaewpuang_2011,Kantere_Dash_2011}. However, a spot service would only offer limited spare resources in practice. As such, a predefined number of homogeneous processors/cores or virtual machines have been widely used to constrain the amount of spot resources \cite{Mazzucco_Dumas_2011,Xu_Li_2013}. Even without specification, the spot service models like $k$ parallel $M/M/1$ queues \cite{Abhishek_Kash_2012} have also implied limited resources ($k$ VM instances in this case) in the Cloud spot market. Similarly, this type of studies did not consider the influence of available resources on spot prices in their modeling work.

(2) The second type of studies focuses on the relationship between spot price and demand/resource. 

\textit{\textbf{Price-Demand model-relevant studies:}}

\begin{figure}[h!t]
\centering
\includegraphics[width=7.3cm]{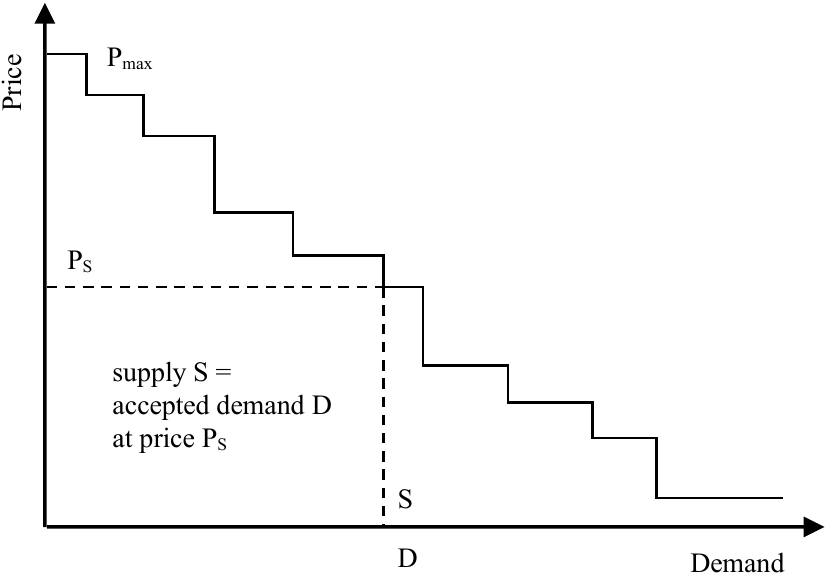}
\caption{Sample demand curve (an adapted version from \cite{Wang_Qi_2013}). The horizontal axis essentially scales accepted demands in the original study.}
\label{fig_DemandCurve}
\end{figure}

Kantere et al.~\cite{Kantere_Dash_2011} modeled the price-demand dependency as second order differential equations with constant parameters, and they claimed that the involved constant parameters could be estimated by using price-demand data sets to perform curve fitting. 
As an important economics concept, the stair-shape demand curve has also been used to represent the relationship between spot price and quantitative demand. For example, Wang et al.~\cite{Wang_Qi_2013} employed a demand curve to facilitate their time-average revenue maximization study. The demand curve was supposed to be maintained by sorting the requests according to their bids in a descending order, as shown in Figure~\ref{fig_DemandCurve}. An interesting feature of this work is that the supply $S$ is defined as the accepted demand $D$ at a particular time slot, which partially emphasizes the relationship between demand and resource of a spot service.

\textit{\textbf{Price-Resource model-relevant studies:}}

The connection between spot price and resource has usually been reflected and/or modeled by using service availability. For example, the lifetimes of spot resources were modeled by building a Markov Chain with edges being the probability of hourly-interval price transitions \cite{Chohan_Castillo_2010}, which is essentially an availability model; on the other hand, the rate of events that terminate application runs due to resource unavailability was considered to follow an exponential distribution \cite{Jangjaimon_Tzeng_2013}, which essentially indicates an unavailability model.

(3) The third type of studies considers the competitions among market participants from the perspective of economics. 

\begin{figure}[h!t]
\centering
\includegraphics[width=7cm]{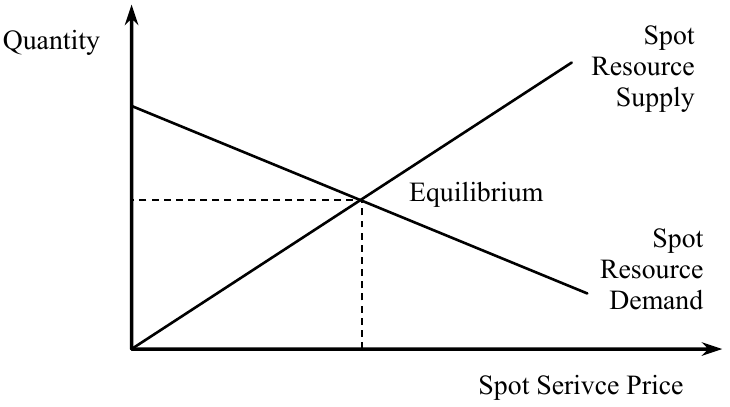}
\caption{Cloud spot pricing from the perspective of economics (originally appears in \protect\cite{Li_Zhang_2015}).}
\label{fig_EconomicsModel}
\end{figure}

In these studies, Cloud spot pricing was treated as various auctions and games to reach some equilibrium. For example, the Prisoner Dilemma game and the Generalized Nash Equilibrium (GNE) game was employed to formulate the conflicts between a provider and its consumers \cite{Di_Valerio_2013,Karunakaran_Sundarraj_2013}. The games and auctions can intuitively explain the influences of spot price and demand/resource on each other, as shown in Figure \ref{fig_EconomicsModel}. However, the equilibrium tends to show a static relationship between demand and resource instead of reflecting their dynamic interactions along the time goes by.

Overall, despite the discussions in common sense, to the best of our knowledge, little work has focused on the interactions between demand and resource in the Cloud spot market. Our work tries to reveal the invisible demand-resource information through a Predator-Prey model.

\section{\uppercase{Using the Predator-Prey Model to Explain Spot Price Variation}}
\label{sec:ppmodel}
Recall that spot price may fluctuate in real time driven by the wax and wane of demands and resources. Given the de facto Cloud spot market, however, we have little knowledge about the changes in demands and resources except for the most recent 90-day price trace disclosed by Amazon \cite{Amazon_2015}. To better understand the market-driven mechanism for Cloud spot service, it would be significantly helpful if the backend information behind spot prices is also visible.

Since it is impossible for us to capture the real-time demands and resources in the market, we decided to use simulation to visualize their changes. Considering that Amazon tends to hold a period of time between different price points \cite{Wee_2011}, we regard Cloud spot service as a discrete-time system \cite{Astrom_Murray_2008}. As such, if viewing the interaction between demand and resource as the relationship between predator and prey by analogy with the two-species ecological system, then it would be natural to employ a Predator-Prey model to realize the simulation.

It is notable that, for investigating ecological systems, the birth and/or death rates are key components in any form of Predator-Prey model \cite{Berryman_1992}. Therefore, we start from determining the birth and death rates of spot service demand and resource before building the model.

\subsection{Birth Rates of Demand and Resource}

Inspired by the explanations in \cite{Xu_Li_2013}, we define the ``birth" of demands as new request arrivals, which is expressed as a Poisson process with rate $f(p)$, and the birth rate $f(p)$ represents the amount of spot resources requested per unit time; while defining the ``birth" of resources as the leave of satisfied requests and the release of available resources, which also follows a Poisson process with rate $g(p)$, and the birth rate $g(p)$ represents the amount of spot resources released per unit time.

Mainly following the assumptions in the previous work \cite{Xu_Li_2013}, we treat the assumed demand arrival and departure rate functions as demand and resource birth rate functions respectively for the potential Predator-Prey model, as shown in Equation (\ref{eq_DemandBirthRate}) and (\ref{eq_ResourceBirthRate}). 

\setlength{\abovedisplayskip}{-6pt} \setlength{\abovedisplayshortskip}{-6pt}

\begin{equation}
\label{eq_DemandBirthRate}
  f(p)=k\times\sqrt[b]{1-p^a}~~~~ (k>0, a>1, b>1)
\end{equation}

\begin{equation}
\label{eq_ResourceBirthRate}
  g(p)=k\times(1-\sqrt[b]{1-p^a})~~~ (k>0, a>1, b>1)
\end{equation}

\begin{figure}[h!t]
  \centering
  \subfigure[Demand birth rate function: {$f(p)=k\times\sqrt[b]{1-p^a}$, $k=5$, $a=3$, $b=3$}.]{
    \label{fig:subfigBirthRates:demand} 
    \makebox[7cm][c]{\includegraphics[width=7cm]{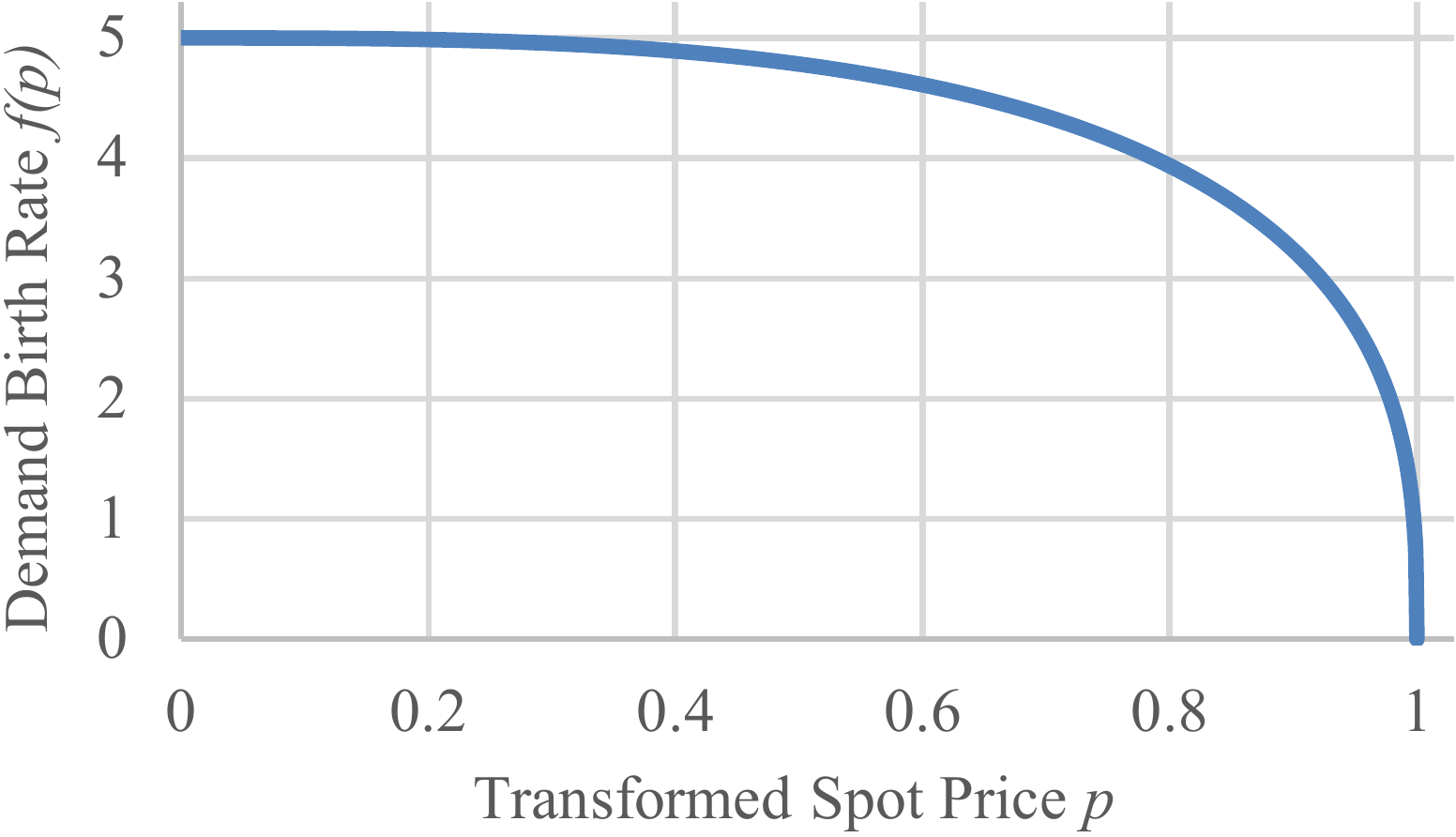}}}
\bigskip
  \subfigure[Resource birth rate function: {$g(p)=k\times(1-\sqrt[b]{1-p^a})$, $k=5$, $a=3$, $b=3$}.]{
    \label{fig:subfigBirthRates:resource} 
    \makebox[7cm][c]{\includegraphics[width=7cm]{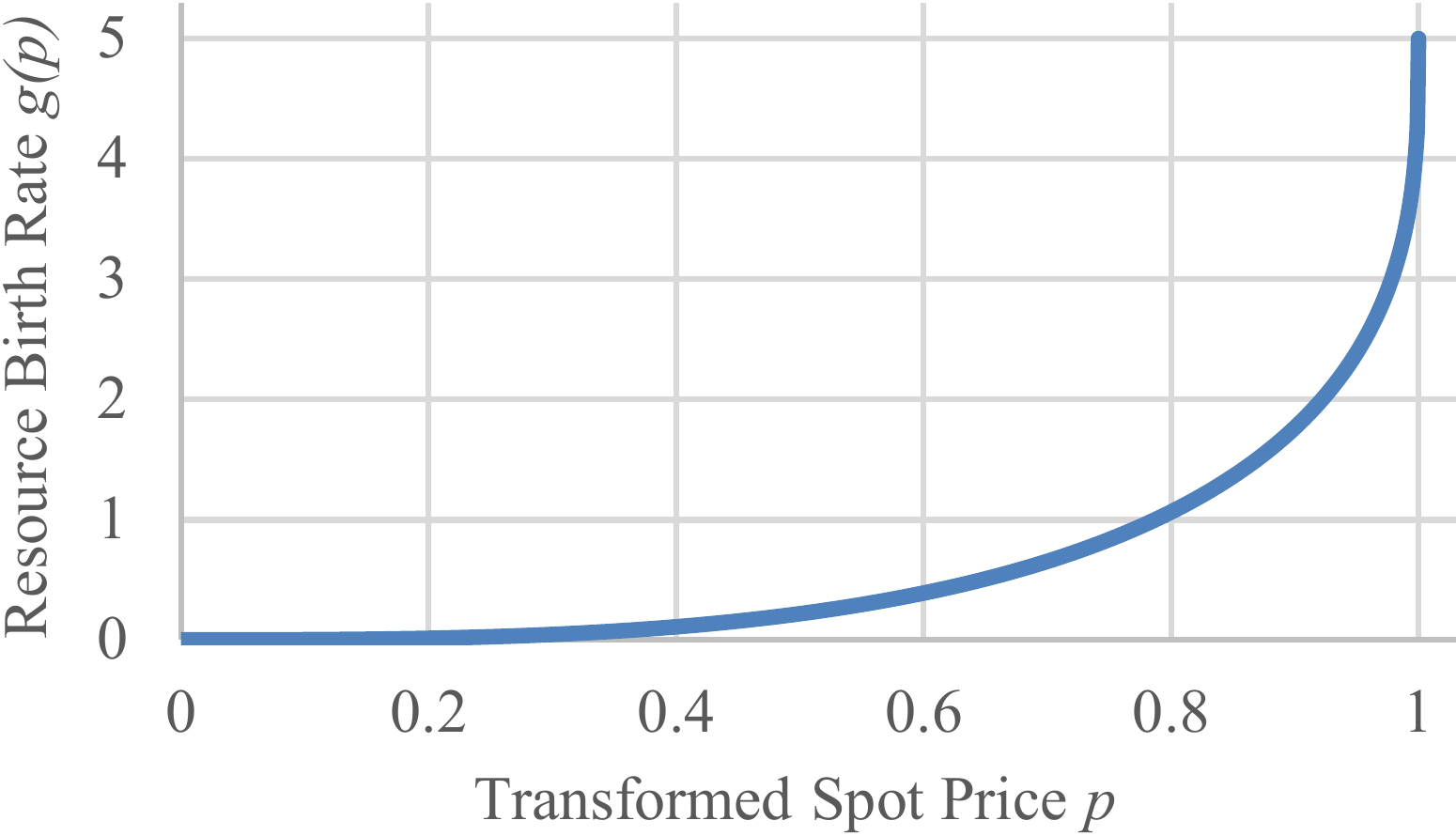}}}
  \caption{Demand and Resource birth rate functions in the Cloud spot market (an adapted version from \cite{Xu_Li_2013}).}
  \label{fig:subfigBirthRates} 
\end{figure}

When it comes to setting values of the parameters, we reuse the example value 5 for $k$ \cite{Xu_Li_2013} while resetting $a$ and $b$ to be 3. In particular, the value of $k$ constrains the ceiling amount of newborn demands or resources, while setting $a$ and $b$ to be 3 can relax the transformed spot price $p$ over an interval wider than $[0, 1]$ if necessary. Note that, to match the birth rate functions, we transform original spot prices by dividing them by their corresponding fixed (on-demand) price. Take Amazon's spot service for example, since spot prices of a particular instance type could be unexpectedly higher than the fixed price \cite{Wee_2011}, some high spot prices would be greater than 1 after transformation, as demonstrated in the middle column of Table \ref{tbl>standardSpotPrice}. In this case, the greater-than-one prices imply a clear discouragement to spot resource employment. In addition, without loss of generality, the bigger value of $a$ and $b$ can make the birth rate functions better align with ``the common psychology" emphasized in \cite{Xu_Li_2013}, as shown in Figure~\ref{fig:subfigBirthRates}: consumers would quickly lose financial incentives to use spot service when its price is approaching the fixed price, while their demands may not be sensitive to the price variation when spot service is far cheaper than the on-demand option.

Furthermore, this transformation is more rational than standardizing spot prices into the interval $[0,1]$, because the standardization will lose the comparability between spot prices and their corresponding fixed price. To reduce the noise of the birth rate functions when the transformed price $p$ is higher than 1, we further round $p$ to 1 if $p>1$ (cf.~Table \ref{tbl>standardSpotPrice}).


\begin{table}[h!t]\footnotesize
\renewcommand{\arraystretch}{1.3}
\centering
\caption{\label{tbl>standardSpotPrice}Transforming Spot Prices of Amazon's Instance Type M3.large}
\begin{tabular}{|>{\raggedright}p{2.5cm} | >{\raggedright\arraybackslash}p{2cm}|l|}
\hline

\textbf{Original Spot Price $p_o$ (USD \$)\textsuperscript{*}} & \textbf{Transformed Spot Price $p$} & \textbf{Rounded $p$}\\
\hline
0.043 & 0.279 &  0.279\\
\hline
0.1 & 0.649 & 0.649\\
\hline
0.14 & 0.909 & 0.909\\
\hline
0.15 & 0.974 & 0.974\\
\hline
0.228 & 1.481 & 1\\
\hline
0.5 & 3.247 & 1\\
\hline
\multicolumn{3}{|l|}{The fixed price of instance type m3.large is \$0.154.}\\
\hline
\multicolumn{3}{>{\raggedright\arraybackslash}p{7cm}}{\textsuperscript{*} The data are from the same price trace as illustrated in Figure~\ref{fig_SpotPriceHistory4P-P}.}\\
\end{tabular}
\end{table}

\subsection{Death Rates of Demand and Resource}

The ``death" of both demand and resource reflects the consumption of Cloud spot service. We define that resources are ``dead" as soon as they are being consumed; and demands are ``dead" as soon as they are being serviced, without waiting for their satisfied moment. 

To determine the demand death rate $\alpha$ and resource death rate $\beta$, we resort to three intuitive assumptions. 
\begin{enumerate}
\setlength{\itemsep}{8pt}
\renewcommand{\labelenumi}{\it{(\theenumi)}}
    \item	We consider 80\% as the death rate for resource if there are more than acceptable amount of demands, i.e., $\beta=0.8$. In fact, due to the possible risks of SLA violation and the inevitable maintenance, Cloud providers would not be interested in a resource utilization that approaches 100\% \cite{Puschel_Borissov_2007}. On the contrary, an average resource utilization of 80\% has been widely considered to be optimal \cite{Puschel_Borissov_2007,Wescott_2013}.
    \item	If the demands are not enough to saturate the optimal service capacity, the amount of dead resource would be equal to the amount of dead demand. In other words, the resource death rate would be less than 80\% in this case.
    \item	Inspired by the Pareto distributions (80-20 rule) \cite{Newman_2005}, we also set 80\% as the death rate for demand no matter whether or not the demands are beyond the optimal service capacity, i.e., $\alpha=0.8$. To unify both situations, we suppose that some out-of-capacity demands would eventually give up employing the spot service, and thus they can also be considered to be dead although without being serviced. In this case, it is still possible to assume only 20\% of demands left for rebidding for the spot service in the next round. 
\end{enumerate} 

In summary, we set the death rate to be 80\% for both spot demand and resource.

\subsection{Predator-Prey Model of Demand and Resource}
\begin{figure}[h!t]
\centering
\includegraphics[width=7cm]{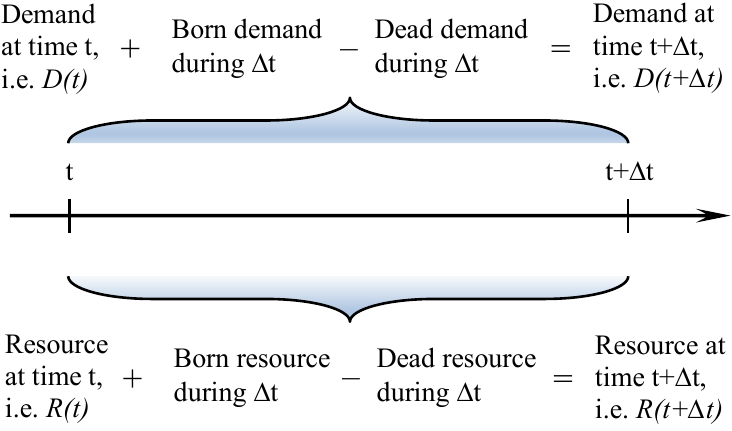}
\caption{Using a timeline to design the logic behind a Predator-Prey model for spot demand and resource.}
\label{fig_ppModel}
\end{figure}

Based on a timeline of spot price variations, we design the intuitive logic behind a Predator-Prey model for recursively exhibiting the amount of demand and resource, as shown in Figure \ref{fig_ppModel}. In detail, we use $D(t)$ and $R(t)$ to refer to the amounts of residual demand and residual resource respectively at time $t$; while $D(t+\Delta t)$ and $R(t+\Delta t)$ respectively represent the amounts of new demand and resource after a period of time $\Delta t$ since $t$. In particular, $\Delta t$ indicates the time span between two price-adjustment points. Taking Amazon as an example, interestingly, its spot service's prices used to be adjusted hourly (i.e.~$\Delta t=1$ hour) \cite{Wee_2011}, while the spot price change frequency becomes multiple times per hour recently (i.e.~$\Delta t<1$ hour) \cite{Guo_Chen_2015}. Note that here we consider the Predator-Prey model by following the logistic thinking of ``principle of population" \cite{Berryman_1992} instead of reusing the difference equations that require interaction terms \cite{Astrom_Murray_2008}. In other words, we claim that the interaction between demand and resource has been reflected by using their both birth and death rates.

By further specifying the logic items with the predefined birth and death rates, we define the Predator-Prey Model of demand and resource in the Cloud spot market, as shown in Equation (\ref{eq_Predator-Prey}).

\setlength{\abovedisplayskip}{0pt} \setlength{\abovedisplayshortskip}{0pt}

\begin{equation}
\label{eq_Predator-Prey}
  \left\{
    \begin{array}{l}
			\begin{aligned}
				D(t+\Delta t)&=D(t) + f(p)\times\Delta t\\
							 &~~~ - D(t)\times\alpha\Delta t\\
			\end{aligned}\\
         ~~\\    
			\begin{aligned}    
         	R(t+\Delta t)&=R(t) + g(p)\times\Delta t\\
							 &~~~ - \min(D(t)\times\alpha\Delta t,R(t)\times\beta\Delta t)\\
			\end{aligned}\\
     \end{array}
   \right.
\end{equation}

To keep a consistent order of magnitude with the predefined values for those birth rate functions (cf.~Fig.~\ref{fig:subfigBirthRates}), we set the initial amounts of spot demand and resource both to 5 at the starting point, i.e., $D(0)=5$ and $R(0)=5$. Furthermore, we assume that the price adjustment happens every unit time interval, and the unit time intervals have uniformly regular sizes, i.e.~$\Delta t=1$ without necessarily considering the unit representation. As such, the impact of time interval $\Delta t$ can be ignored when calculating both the birth and the death of spot demands and resources.

Overall, an example setting for initializing the parameters of this model is specified in Equation (\ref{eq_Predator-Prey-parameters}).

\begin{equation}
\label{eq_Predator-Prey-parameters}
  \left\{
    \begin{array}{l}
         f(p)=5\times\sqrt[3]{1-p^3}\\
         ~~\\
         g(p)=5\times(1-\sqrt[3]{1-p^3})\\
         ~~\\
         D(0)=R(0)=5,~ \alpha=\beta=0.8,~\Delta t=1
     \end{array}
   \right.
\end{equation}

\section{\uppercase{Simulation using Amazon's Spot Price Trace}}
\label{sec:simulation}

\begin{algorithm}[h!t]\footnotesize
\caption{Demand-Resource Interaction Simulation}
\label{PPAlgorithm}
        \textbf{Input:} Array of historical spot prices $P=(p_1, p_2, p_3,...,p_t)$, initial demand amount $d_0$, initial resource amount $r_0$, birth rate factors $a, b, k$, demand death rate $\alpha$, resource death rate $\beta$. \\
        \textbf{Output:} Array of demand amount $D=(d_1, d_2,..., d_t)$, array of resource amount $R=(r_1, r_2,..., r_t)$.\\
\begin{algorithmic}[1]
\Function{Dbr}{$p$} \Comment{Demand birth rate at price p}
  \State $f\gets k\times (1-p^a)^{1/b}$ 
  \State \Return $f$
\EndFunction

\Function{Rbr}{$p$} \Comment{Resource birth rate at price p}
  \State $g \gets k\times [1-(1-p^a)^{1/b}]$ 
  \State \Return $g$
\EndFunction

\State $d_t \gets d_0$ \Comment{Initial current demand amount at time t}
\State $r_t \gets r_0$ \Comment{Initial current resource amount at time t}
\State $d_{t+1} \gets 0$ \Comment{Initial new demand amount at time t+1}
\State $r_{t+1} \gets 0$ \Comment{Initial new resource amount at time t+1}
\State $D \gets \emptyset \cup d_0$ \Comment{Initial array of demand amount}
\State $R \gets \emptyset \cup r_0$ \Comment{Initial array of resource amount}
\For{$j=1,2,3,...,t$ }
\State $d_{t+1} \gets d_t + Dbr(p_j)- \alpha\times d_t$
\State $r_{t+1} \gets r_t + Rbr(p_j)- \min\{\alpha\times d_t,\beta\times r_t\}$
\State $D \gets D \cup d_{t+1}$
\State $R \gets R \cup r_{t+1}$
\State $d_t \gets d_{t+1}$
\State $r_t \gets r_{t+1}$
\EndFor
\State \Return $D,R$
\end{algorithmic}
\end{algorithm}

\begin{figure}[h!t]
\centering
\includegraphics[width=7.3cm]{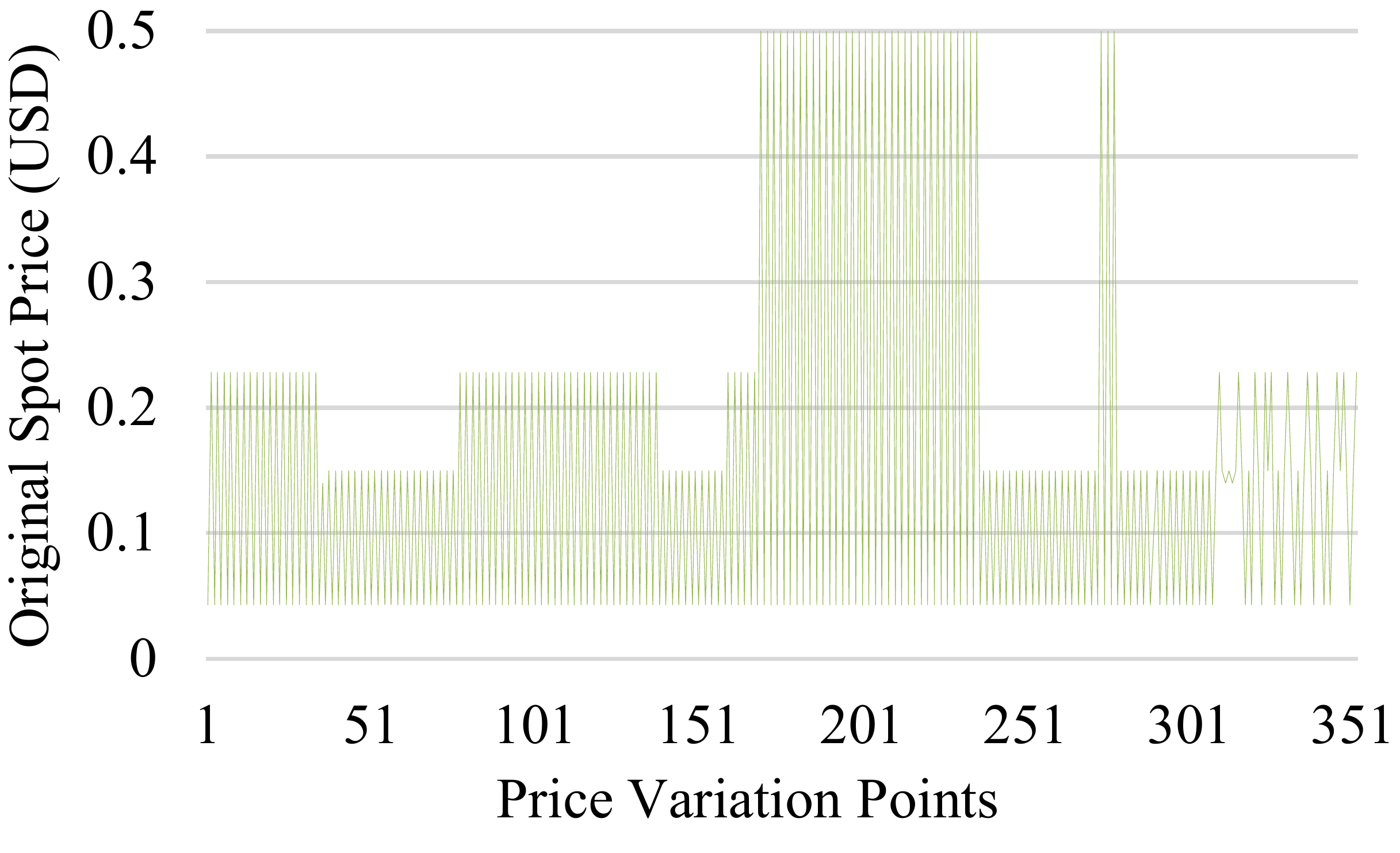}
\caption{Amazon's spot price variation trace between 2015-03-12 00:01:10 and 2015-03-16 09:23:56 (instance type: m3.large, OS type: Linux/UNIX, zone: us-east-1b).}
\label{fig_SpotPriceHistory4P-P}
\end{figure}

Given particular initialization settings (e.g.,~Equation (\ref{eq_Predator-Prey-parameters})), this Predator-Prey model can generate demand and resource simulations corresponding to spot price traces. To facilitate simulation, we implement the Predator-Prey model into executable codes, as specified
in Algorithm \ref{PPAlgorithm}. This straightforward algorithm also shows
that replicating our study would not be
difficult.
When it comes to the spot price trace, we use Amazon's Command Line Interface (CLI) tool ec2-describe-spot-price-history \cite{Amazon_2015_tool} to collect historical spot prices of the instance type m3.large whose price has relatively frequent fluctuations at the time of writing. For the purpose of conciseness, we only select a typical piece of data (spot price records between 2015-03-12 00:01:10 and 2015-03-16 09:23:56) in the collected full trace\footnote{The complete spot price trace with 24000 records of spot instance type m3.large has been shared online: \url{https://docs.google.com/spreadsheets/d/1mBUItaGwsa44J5_DHLS0H_mh9QLKjOQPTl7Oq7aiuAc/}}, as illustrated in Figure~\ref{fig_SpotPriceHistory4P-P}. Note that not all the historical spot prices can be used to generate reasonable simulations. Only frequently oscillating price traces can fit in our Predator-Prey model. In fact, a piece of flat price trajectory might indicate a lack of demands during that time period, and therefore leading to few demand-resource interactions. Such a scenario cannot employ any Predator-Prey model, because it deviates from the natural Predator-Prey rules.

\begin{figure}[h!t]
\centering
\includegraphics[width=7.3cm]{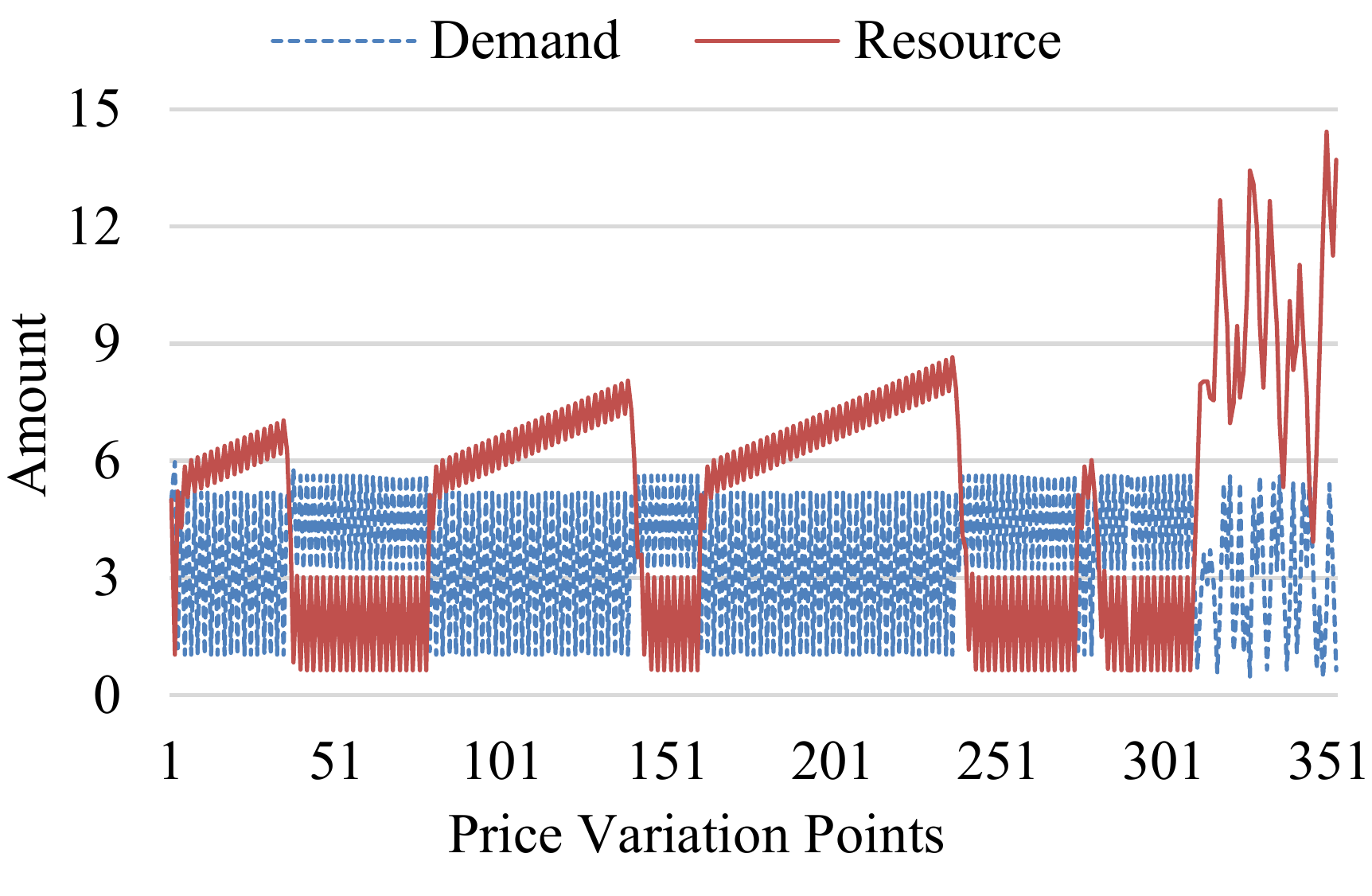}
\caption{Demand and Resource simulation corresponding to Amazon's spot price variation trace between 2015-03-12 00:01:10 and 2015-03-16 09:23:56 (instance type: m3.large, OS type: Linux/UNIX, zone: us-east-1b).}
\label{fig_Simulation4P-P}
\end{figure}


Following the consecutive time series in the selected price trace, the sequential amounts of demand and resource can be calculated along with the changing spot prices, as plotted in Figure~\ref{fig_Simulation4P-P}. As mentioned previously, we use the rounded transformed prices for the calculations (cf.~Table \ref{tbl>standardSpotPrice}). It is clear that, although we have made simplifying assumptions for building this Predator-Prey model, the simulation here can still help reveal basic information behind spot prices. For example:
\begin{itemize}
 \item{Consumer demands remain at a low level when the spot service is generally expensive.}
 \item{Spare resources remain at a low level when the spot service is generally cheap.}
 \item{Spot resources would gradually be released rather than a burst when the spot service is becoming expensive.}
 \item{In contrast, the spot service capacity could quickly be saturated by attracting demands at low spot prices.}
\end{itemize}
 
The first two simulation findings are aligned with our common sense, while the others require further validation in practice. In particular, the frequent sharp drops of resources revealed by the fourth finding might indicate the regular herd behaviors of spot demands. Originally, herd behavior refers to a typical phenomenon when a group of individuals act collectively without centralized direction, and it could happen among animals as well as humans \cite{Braha_2012}. By analogy, the herd behavior of spot demands could result from the same behavior of users when bidding for cheap Cloud spot resources. Although it is difficult to make validation at this current stage due to the lack of practical data, we can treat this explanation as a hypothesis to be tested in the future.

\section{\uppercase{Conclusions and Future Work}}
\label{sec:conclusion}
Among the three typical pricing schemes in the de facto Cloud market, spot pricing has been widely accepted as the most resource-efficient strategy for Cloud providers and the most cost-effective option for Cloud consumers. Nevertheless, the spot pricing scheme seems not to be popular yet for trading Cloud resources, because the market-driven mechanism for pricing spot service would be complicated both for providers to implement and for consumers to understand. Despite limited historical spot prices disclosed by Amazon, the existing spot market activities are generally invisible especially for Cloud consumers. 

To help better understand the operations in the Cloud spot market, we developed a Predator-Prey model to visualize the potential demand-resource interactions based on the available spot price traces. The simulation study has revealed some basic information behind spot price variations, and also proposed a hypothesis guiding us in future validation. Overall, our work essentially shows that utilizing a Predator-Prey model could be a promising approach to reversely engineer spot market activities. 

However, there is still a lack of practical data to validate our simulation findings. Such a limitation drives our future work along two directions. On the one hand, we will gradually improve our Predator-Prey model by employing relatively solid assumptions. For example, more sophisticated mathematical models can be used to represent the death rates of demand and resource. On the other hand, we will try to extract useful demand and resource data from workload traces, and conduct workload characterization to verify the findings of this study.

\section*{\uppercase{Acknowledgements}}

\noindent 
This work is funded in part by the Swedish Research Council (VR) under contract number C0590801 for the project Cloud Control.
We are also members of the Lund Center for Control of Complex Engineering Systems (LCCC) funded by the Swedish Research Council (VR) and the Excellence Center Link{\"{o}}ping - Lund in Information Technology (ELLIIT).

\bibliographystyle{apalike}
{\small
\bibliography{Closer_Ref}}

\end{document}